\documentclass[pra,aps,twocolumn,floatfix]{revtex4}
\usepackage{graphicx,graphics,psfrag,amsmath,calc}
\usepackage{epsfig}
\usepackage{color, comment}
\usepackage{mathbbol}
\usepackage{float}
\usepackage{placeins}
\usepackage{subfigure}

\begin{document}

\title{
Chern number spectrum of ultra-cold fermions in optical lattices \\
tuned independently via artificial magnetic, Zeeman and spin-orbit fields}

\author{Man Hon Yau and C. A. R. S\'a de Melo}

\affiliation{
School of Physics, Georgia Institute of Technology, 
Atlanta, 30332, USA
}

\date{\today}

\begin{abstract}
We discuss the Chern number spectrum of ultra-cold fermions in square 
optical lattices as a function artificial magnetic, Zeeman and spin-orbit 
fields that can be tuned independently. 
We show the existence of topological quantum phase transitions 
induced by Zeeman and spin-orbit fields, where the total number and 
chirality of edge states change for fixed magnetic flux ratio, 
thus leading to topological-insulator phases which are different
from those found at zero Zeeman and spin-orbit fields. 
We construct phase diagrams of chemical potential versus Zeeman field 
or spin-orbit coupling and characterize all insulating phases by their
topological invariants. Lastly, we obtain a staircase structure in the filling
factor versus chemical potential for various Zeeman and spin-orbit fields, 
showing the existence of incompressible states at rational filling factors 
derived from a generalized Diophantine equation.  
\end{abstract}

\maketitle

%
%

%
%

Ultra-cold fermions loaded in optical lattices have become ideal systems to 
study related electronic phase diagrams and transport
properties, because they provide a clean and well controlled playground 
to change various lattice parameters and external fields at the turn 
of a knob. It is now possible to create 
artificial magnetic fields~\cite{bloch-2013, ketterle-2013} in optical 
lattices that mimic electronic materials exhibiting  
integer~\cite{klitzing-1980} and fractional~\cite{stormer-1982} 
quantum Hall effects. The synthetic magnetic flux values reached are 
sufficiently large to allow for the experimental exploration 
of the intricacies of the Harper's model~\cite{harper-1955} and the Hofstadter 
butterfly~\cite{hofstadter-1976}, such as the experimental determination
of Chern numbers~\cite{bloch-2014}. Furthermore the creation of artificial 
spin-orbit coupling for ultra-cold atoms~\cite{spielman-2011} allows for  
the simulation of electronic materials exhibiting the quantum spin-Hall 
effect~\cite{kane-2005,haldane-2005,zhang-2006}.  

For ultra-cold fermions, artificial magnetic fields allow for 
studies of topological insulators that break time reversal symmetry, 
such as quantum hall systems, while artificial spin-orbit fields allow  
for studies of topological insulators that do not break time reversal symmetry, 
such as quantum spin-Hall systems. Both types of topological insulators are
characterized by Berry curvatures and Chern numbers, which have been 
measured using time of flight techniques~\cite{weitenberg-2016}, 
inspired by theoretical proposals~\cite{indu-2011, lewenstein-2014}, and
using dynamics of the center of mass of the atomic cloud~\cite{esslinger-2014},
also motivated by theoretical work~\cite{cooper-2012, goldman-2013a}.
However, studies of ultra-cold fermions can 
go way beyond the quantum simulation of topological insulators under typical 
condensed matter conditions. The independent tunability of artificial magnetic, 
spin-orbit and Zeeman fields in cold atoms is possible via the combination 
of experimental techniques that produce artificial magnetic fluxes without 
using internal states, such as laser assisted 
tunneling~\cite{bloch-2013, ketterle-2013}, or that produce 
spin-orbit and Zeeman fields using internal states, such 
a Raman processes~\cite{spielman-2011} or 
radio-frequency chips~\cite{spielman-2010}. 

In this paper, we study the interplay of artificial magnetic, 
spin-orbit and Zeeman fields and their effects on topological insulators 
in regimes that cannot be reached in condensed matter physics.
We analyse the Chern number spectrum of fermionic atoms with two
internal states, such as, $^{6}$Li or $^{40}$K, and show that topological
quantum phase transitions are induced by Zeeman and spin-orbit fields 
at constant magnetic flux. Finally, we
construct phase diagrams of chemical potential versus Zeeman field 
or spin-orbit coupling and characterize all insulating phases by their
topological invariants.

{\it Hamiltonian:} 
We begin our investigation by writing the first quantization 
Hamiltonian matrix for ultra-cold fermions in a two-dimensional square 
optical lattice as 
\begin{eqnarray} 
\label{eqn:hamiltonian-matrix}
{\hat{H}} 
= 
\left(
\begin{array}{c c }
\varepsilon_{\uparrow} ( \hat{\bf k}) & -h_x      \\
-h_x            &  \varepsilon_{\downarrow} (\hat {\bf k})
\end{array} 
\right), 
\end{eqnarray}
where 
$
\varepsilon_{\uparrow} (\hat{\bf k}) =
-2t \{\cos[({\hat k}_x - k_{T})a] + \cos[({\hat k}_y  - {\cal A}_y)a]\}
$
corresponds to the spin-up $(\uparrow)$ kinetic energy and 
$
\varepsilon_{\downarrow} (\hat{\bf k}) =
-2t \{\cos[({\hat k}_x + k_{T})a] + \cos[({\hat k}_y  -{\cal A}_y)a]\}
$
corresponds to the spin-down $(\downarrow)$ kinetic energy.
Here, $t$ is the hopping amplitude, $a$ is the lattice spacing, 
$k_T$ is a spin-dependent momentum transfer characterizing 
an artificial unidirectional (one-dimensional) spin-orbit coupling, 
and ${\cal A}_y = eHx/\hbar c$ plays the role of the $y$-component 
of an artificial vector potential, where $H$ is identified as 
a synthetic magnetic field along the $z$-axis. Notice ${\cal A}_y$ 
has dimensions of inverse length. 
Lastly, $h_x$ represents a Zeeman field along the $x$-direction, 
whose physical origin is a Rabi {\it spin-flip} 
term that couples the two internal states of the atom. 
The vector potential ${\cal A}_y$ may be generated by laser assisted 
tunneling~\cite{bloch-2013, ketterle-2013}, while the spin-dependent momentum
transfer $k_T$ (spin-orbit) and Zeeman field $h_x$ may be created via 
counter-propagating Raman beams~\cite{spielman-2011} or via radio-frequency
chips~\cite{spielman-2010}.

{\it Eigenspectrum:}
First, we find the eigenspectrum of the Hamiltonian matrix described in 
Eq.~(\ref{eqn:hamiltonian-matrix}) as a function of magnetic flux through a 
lattice plaquette. We work in  a cylindrical geometry having finite number
$N$ of sites along the $x$-direction and open boundaries, but periodic 
boundary conditions along the $y$-direction. 

In this case, the spin-dependent Harper's matrix 
\begin{eqnarray}
\label{eqn:hamiltonian-matrix-cylinder-geometry}
{\bf H} 
=
\left(
\begin{array}{c c c c c}
{\bf A}_{m-2}  & {\bf B}       &  {\bf 0}      &  {\bf 0}      & {\bf 0}       \\
{\bf B}^*      & {\bf A}_{m-1} &  {\bf B}      &  {\bf 0}      & {\bf 0}       \\
{\bf 0}        & {\bf B}^*     &  {\bf A}_m    & {\bf B}       & {\bf 0}       \\
{\bf 0}        & {\bf 0}       &  {\bf B}^*    & {\bf A}_{m+1} & {\bf B}       \\ 
{\bf 0}        & {\bf 0}       &  {\bf 0}      & {\bf B}^*     & {\bf A}_{m+2} \\
\end{array}
\right)
\end{eqnarray}
has a tridiagonal block structure that couples neighboring sites 
$(m-1, m, m+1)$ along the $x$-direction and takes full advantage of discrete 
translational invariance along the $y$-axis. The matrices
${\bf A}$, ${\bf B}$ and the null matrix ${\bf 0}$ 
consist of $2\times2$ blocks with entries labeled by internal 
states $\vert \uparrow \rangle$ and $\vert \downarrow \rangle$. 
The size of the space labeled by the site index $m$ is $N$, thus 
the total dimension of the matrix ${\bf H}$ in 
Eq.~(\ref{eqn:hamiltonian-matrix-cylinder-geometry})
is $2N \times 2N$. The matrix indexed by position $x = m a$ is   
\begin{eqnarray}
{\bf A}_m 
=
\left(
\begin{array}{cc}
-2 t \cos(k_y a - 2\pi m \alpha) & -h_x \\
- h_x  & -2 t \cos(k_y a - 2\pi m \alpha)
\end{array}
\right),
\nonumber
\end{eqnarray}
where the parameter $\alpha = \Phi/\Phi_0$ represents the ratio
between the magnetic flux though a lattice plaquete $\Phi = H a^2$ and
the flux quantum $\Phi_0 = hc/e$ or the ratio between the 
plaquette area $a^2$ and the square of the magnetic length 
$\ell_{M} = hc/eH$, that is, $\alpha = (a/\ell_M)^2$.
The matrix containing the spin-orbit coupling is 
\begin{eqnarray}
{\bf B}
=
\left(
\begin{array}{cc}
- t e^{-ik_{T}a} & 0 \\
0 & - t e^{ik_{T}a}
\end{array}
\right)
\nonumber
\end{eqnarray}
corresponding to momentum shift by $k_T$ $(-k_T)$ along the $x$-direction for the 
$\vert \uparrow \rangle$ $(\vert \downarrow \rangle)$ state.
Notice that in a cylindrical geometry with open boundaries 
in the $x$-direction, but
periodic boundary condictions along the $y$-direction, $k_y$ is a good 
quantum number, while $k_x$ is not.

%
%

\begin{figure}[tb]
\centering 
\epsfig{file=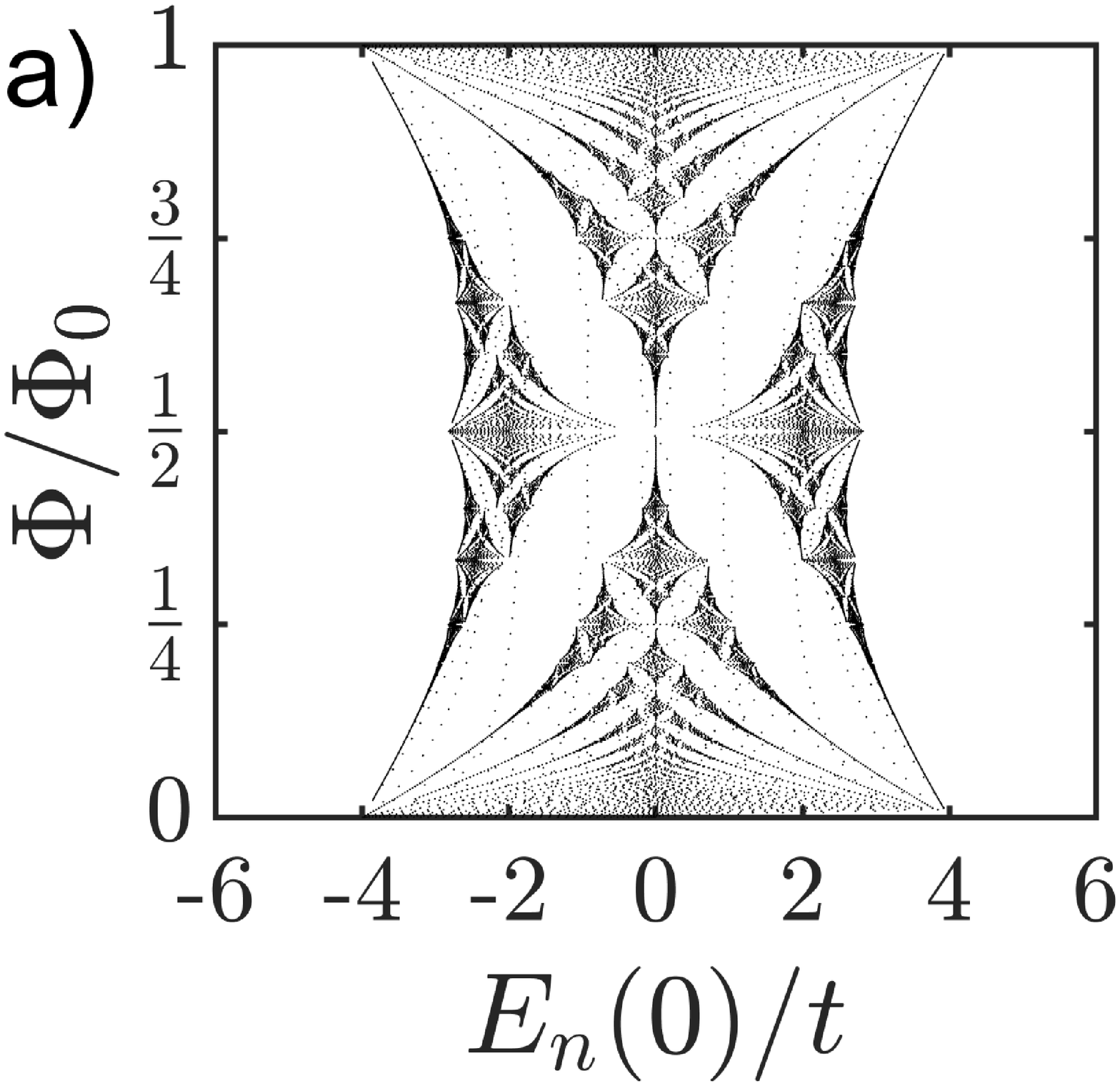,width=0.49 \linewidth}
\epsfig{file=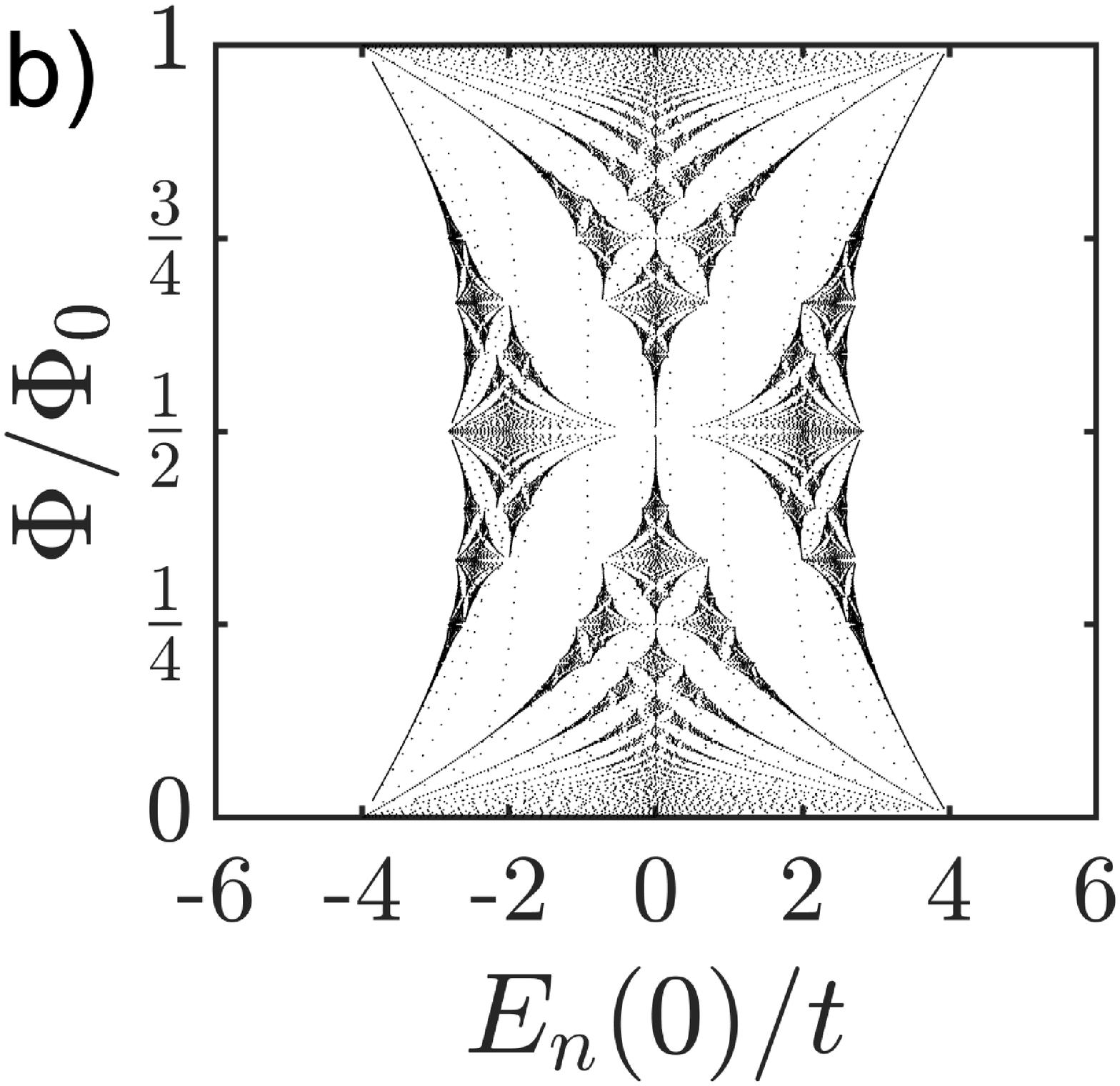,width=0.49 \linewidth}
\vskip 0.2cm
\epsfig{file=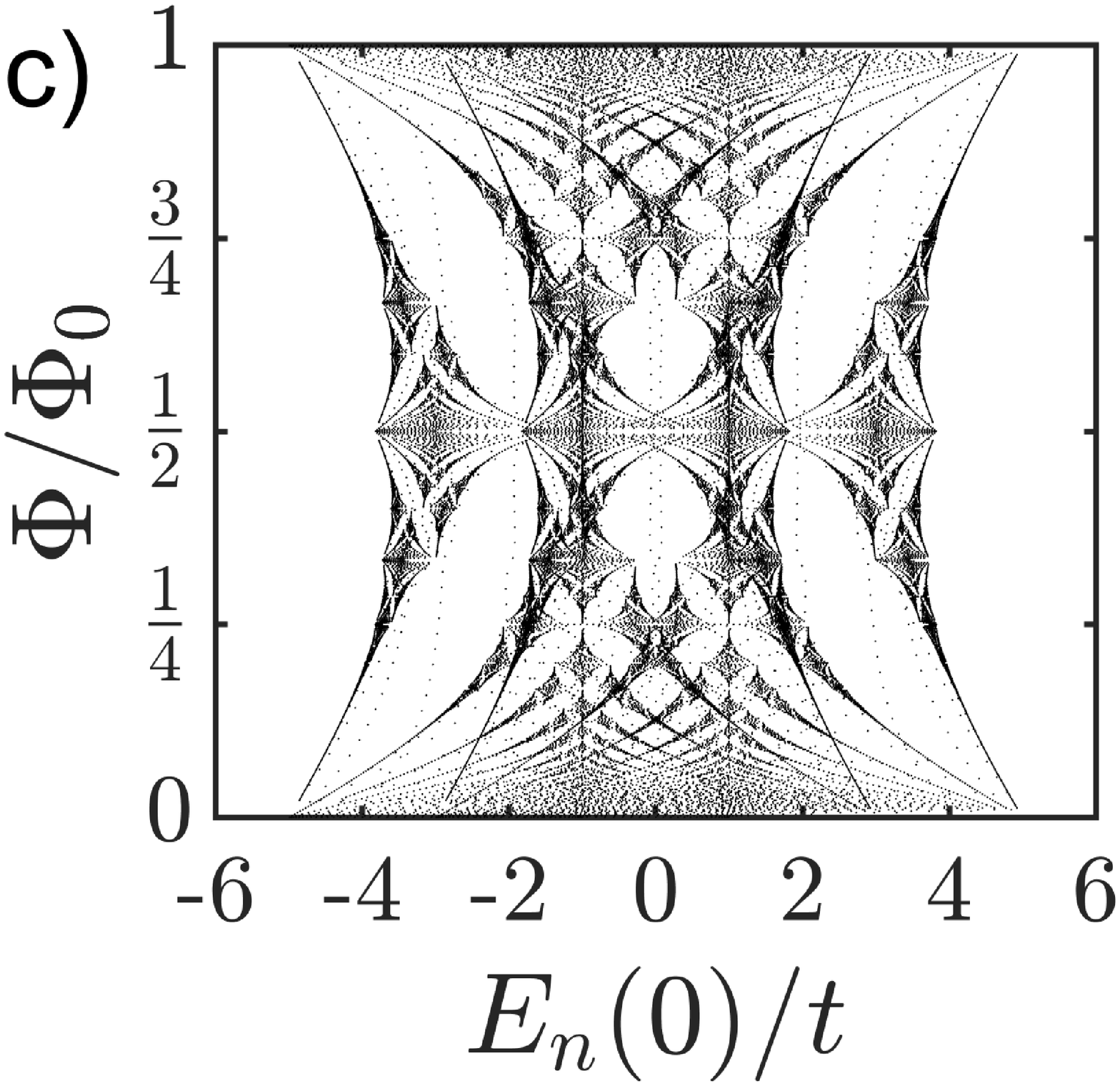,width=0.49 \linewidth}
\epsfig{file=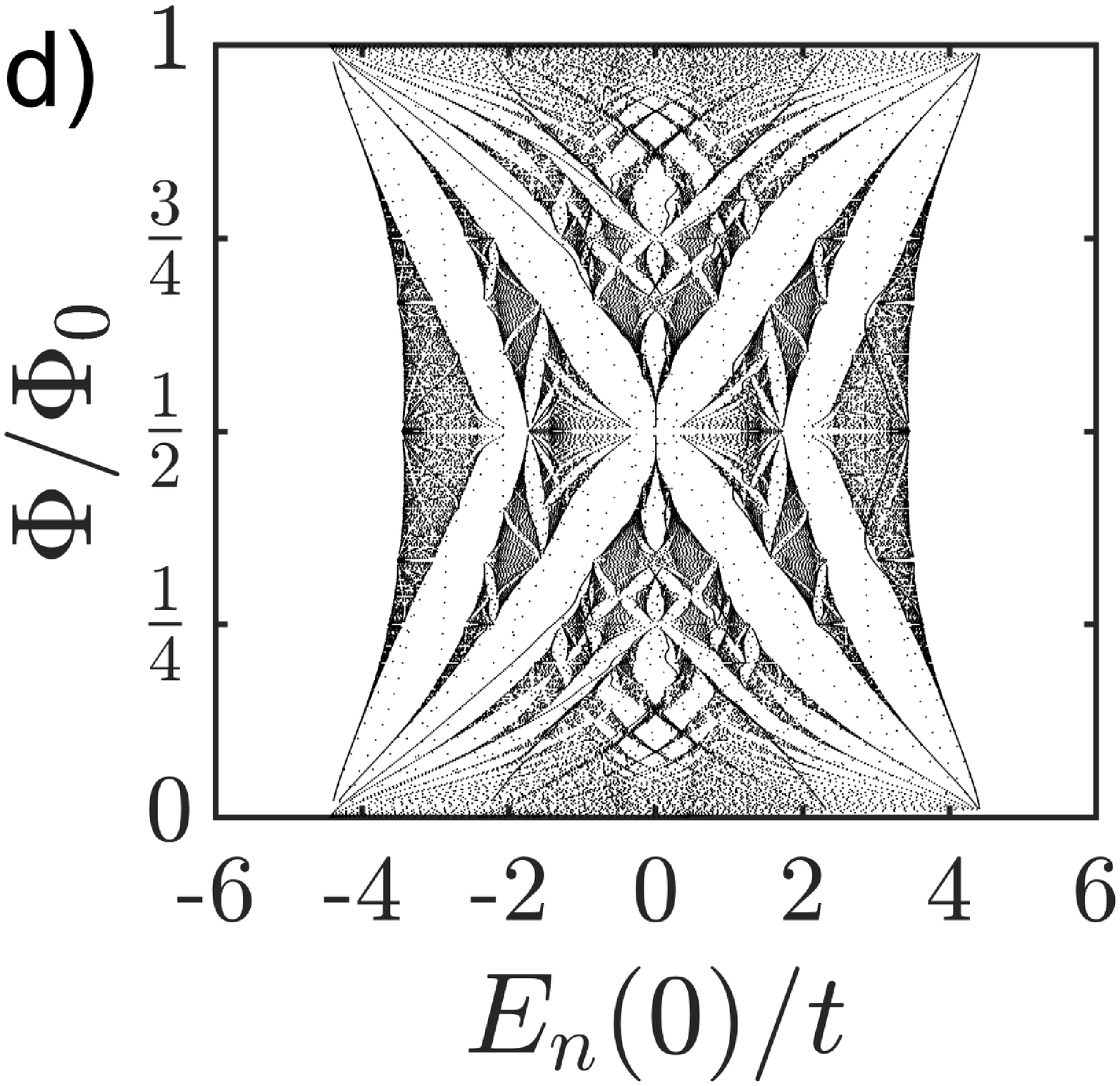,width=0.49 \linewidth}
\caption{ 
\label{fig:one}
Flux ratio $\alpha = \Phi/\Phi_0$ versus energy 
$E_n (k_y = 0)$ for various values of spin-orbit coupling 
parameter $k_T a$ and Zeeman field $h_x/t$. 
The parameters are: 
a) $k_T a = 0$ and $h_x/t = 0$,  b) $k_T a = \pi/4$ and $h_x/t = 0$, 
c) $k_T a = 0$ and $h_x/t = 1$, d) $k_T a = \pi/4 $ and $h_x/t = 1$.
}
\end{figure}

The generalized Hofstadter spectrum can be obtained from the
spin-dependent Harper's matrix defined in 
Eq.~(\ref{eqn:hamiltonian-matrix-cylinder-geometry}). We consider the
dimensions of the optical lattice to be of fifty (50) sites along 
the $x$-direction,
with two spin states per site, but periodic along the $y$-direction. 
The eigenvalues $E_{n}( k_y )$ are labeled by a discrete band 
index $n$ and by 
momentum $k_y$, and are functions of the spin-orbit coupling 
$k_T$, Zeeman
field $h_x$, as well as flux ratio $\alpha = \Phi/\Phi_0$.  
In Fig.~\ref{fig:one}, we show the spectrum of 
$\alpha = \Phi/\Phi_0$ versus zero-momentum energy $E_n (k_y = 0)$ 
for the following cases: 
a) zero spin-orbit $(k_T a = 0)$ and zero Zeeman 
field $(h_x/t = 0)$,
corresponding to the standard {\it butterfly} graph; 
b) spin-orbit coupling $k_T a = \pi/4$ and zero Zeeman 
field field $h_x/t = 0$ producing the same graph as in a) 
due to a spin-gauge symmetry that allows gauging away the spin-orbit 
coupling when $h_x/t = 0$; 
c) zero spin-orbit coupling $(k_T a = 0)$ and Zeeman field
$h_x/t = 1$ producing two Zeeman shifted, but interpenetrating 
{\it butterfly} graphs; d) spin-orbit coupling $k_T a = \pi/4$ 
and Zeeman field $h_x/t = 1$ leading to a richer
{\it butterfly-spider} spectrum.

%

%
%

\begin{figure} [tb]
\centering 
\epsfig{file=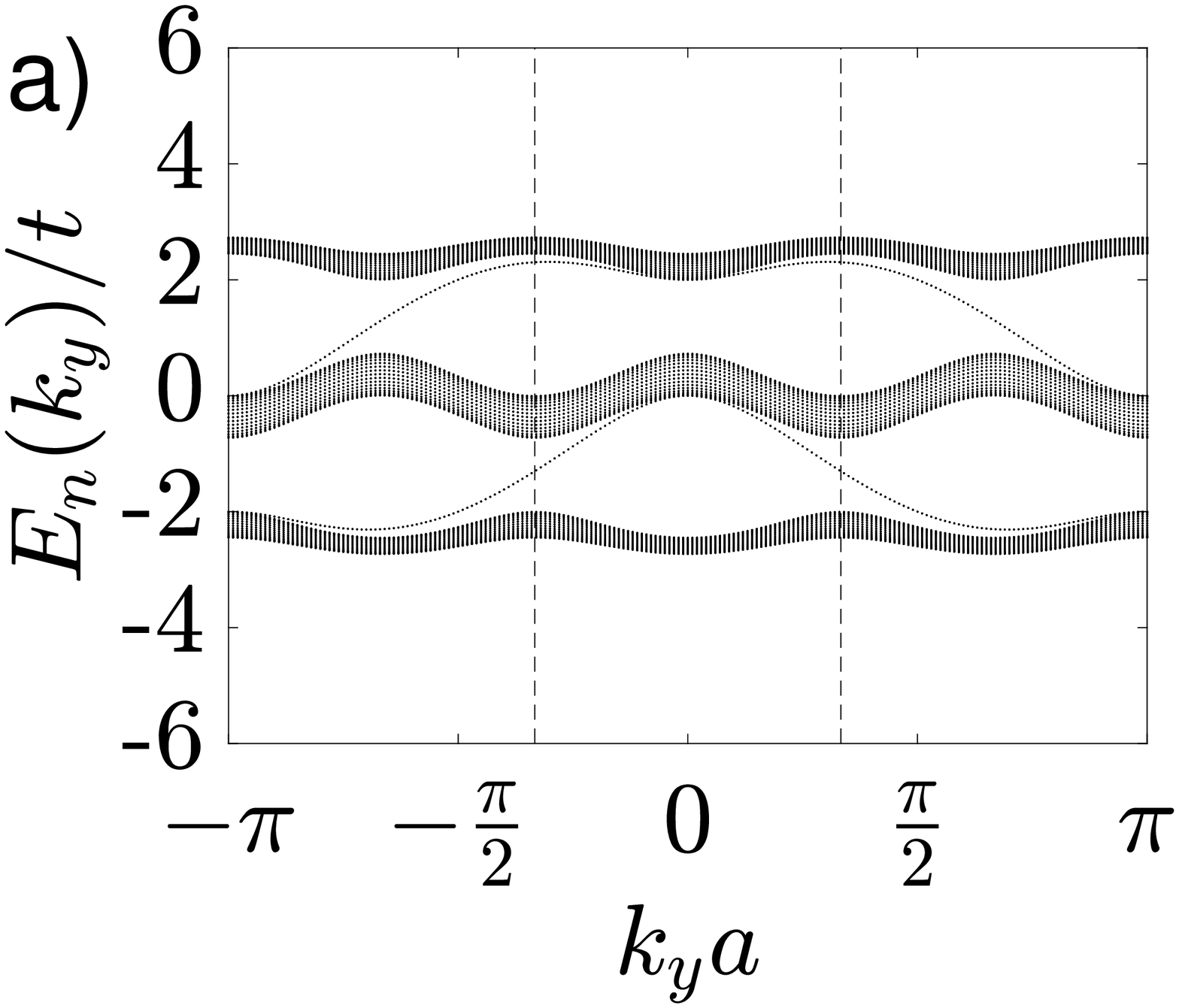,width=0.49 \linewidth}
\epsfig{file=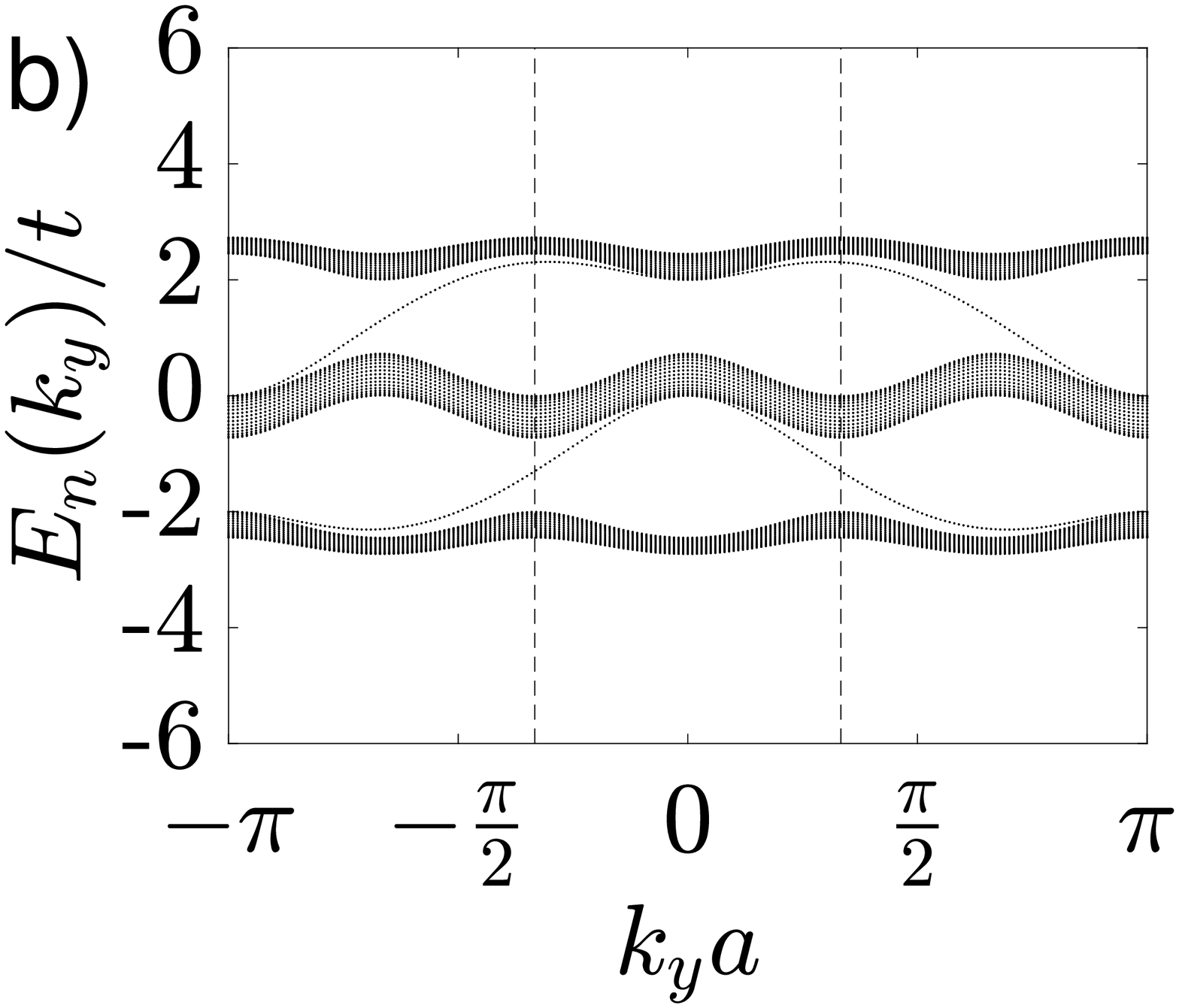,width=0.49 \linewidth}
\vskip 0.2cm
\epsfig{file=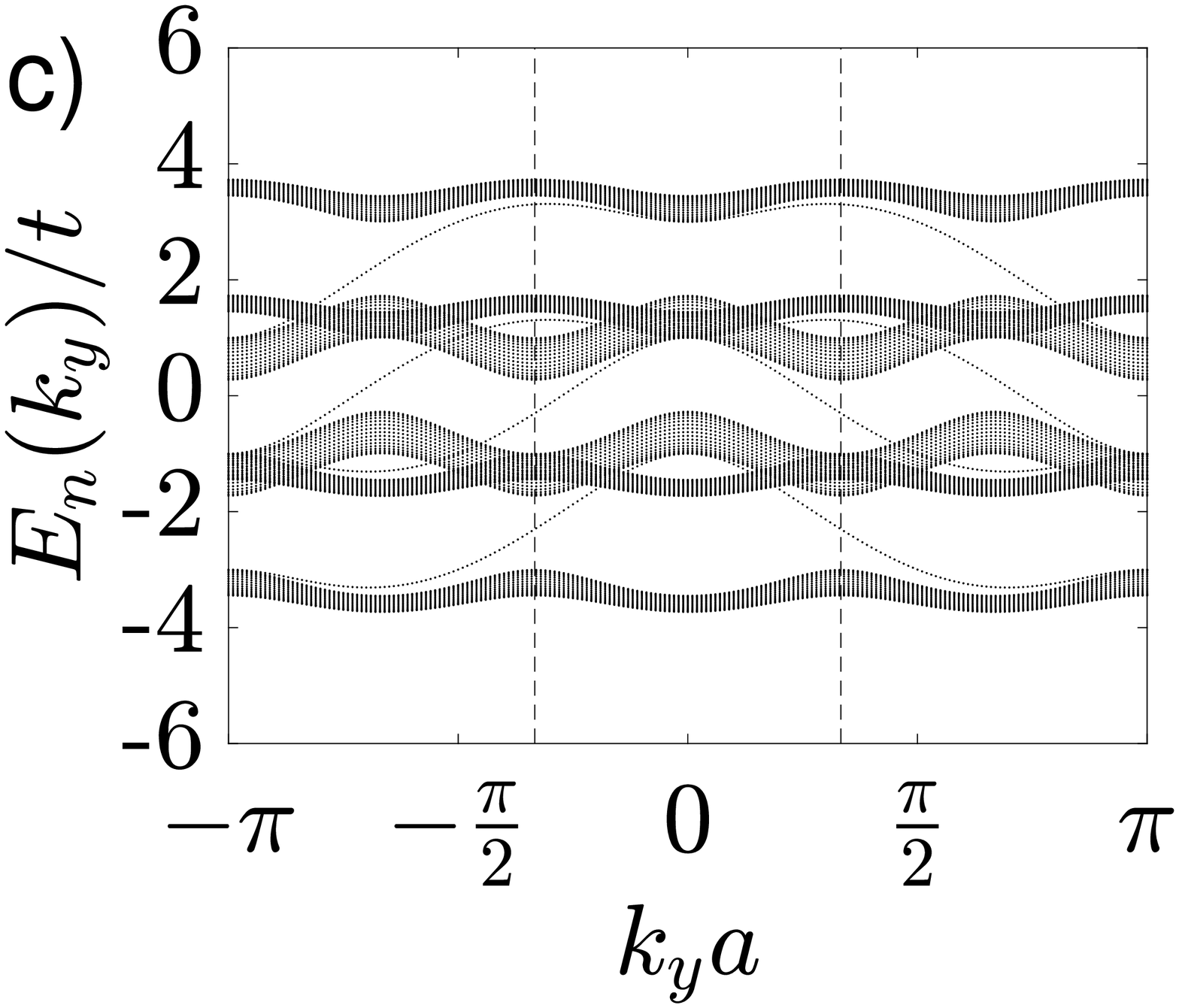,width=0.49 \linewidth}
\epsfig{file=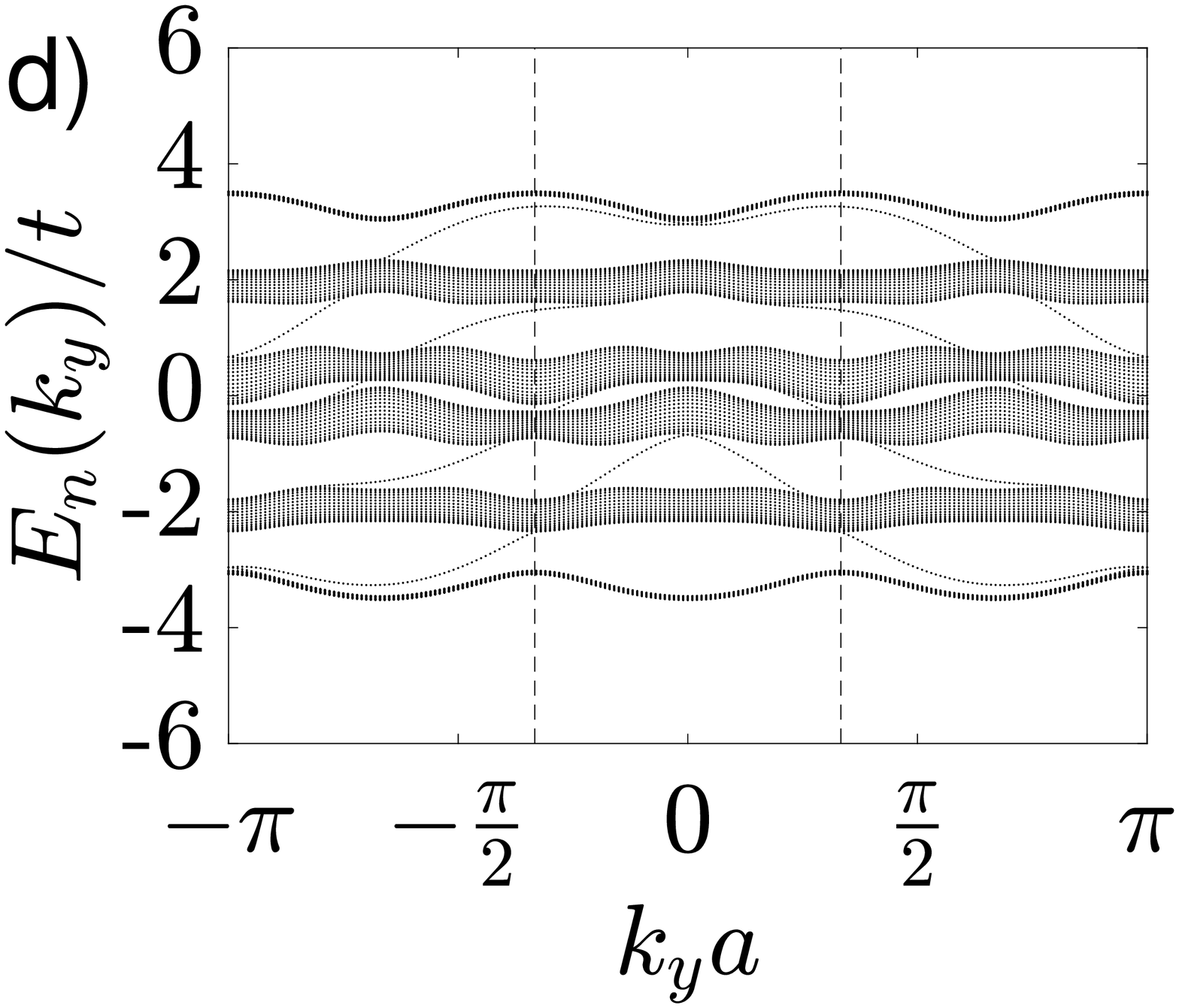,width=0.49 \linewidth}
\caption{ 
\label{fig:two}
Eigenvalues $E_n (k_y)$ of the spin-dependent Harper's matrix
versus $k_y a$ for magnetic flux ratio $\alpha = 1/3$. 
The parameters are: a) $k_T a = 0$ and $h_x/t = 0$,  
b) $k_T a = \pi/4$ and $h_x/t = 0$, 
c) $k_T a = 0$ and $h_x/t = 1$, d) $k_T a  = \pi/4$ and $h_x/t = 1$. 
The vertical dashed lines located at $k_y a = \pm \pi/3$ indicate the 
boundaries of the magnetic Brillouin zone. The bulk 
bands have periodicity $2\pi/3a$, and the 
edge bands have periodicity $2\pi/a$ along the $k_y$ direction.
}
\end{figure}

We discuss next the particular case of flux ratio $\alpha = 1/3$ for the 
cylindrical geometry described above and analyze the eigenvalues 
$E_n (k_y)$ for a few values 
of spin-orbit parameter $k_T$ and Zeeman field $h_x$. 
In Fig.~\ref{fig:two}, we show the cases: 
a) $k_T a = 0$ and $h_x/t = 0$, which contains three sets of 
degenerate bulk bands connected by spin-degenerate edge bands; 
b) $k_T a = \pi/4$ and $h_x/t = 0$, which is identical to 
case a) because of a spin-gauge symmetry that allows gauging away 
the spin-orbit coupling; 
c) $k_T a = 0$ and $h_x/t = 1$ contains six sets 
of bulk bands because spin-degeneracies are lifted, and 
spin-dependent edge bands; 
d) $k_T a = \pi/4$ and $h_x/t = 1$, which contains six sets 
of bulk bands connected by spin-dependent edge states, 
all subjected to simultaneous effects of spin-orbit coupling 
and Zeeman field. Bulk bands in all panels
have momentum space period of $2\pi/3a$, while the edge 
bands have period $2\pi/a$ along the $k_y$ direction.
The periodicity of the bulk states is dictated by the denominator 
$q$ of the rational magnetic flux ratio $\alpha = p/q$, which 
for $\alpha = 1/3$ corresponds to $p = 1$ and $q = 3$. 
In Fig.~\ref{fig:two}, the vertical dashed lines 
indicate the boundaries of the magnetic Brillouin zone at 
$k_y a = \pm \pi/3$. 

{\it Chern numbers:}
To study the Chern number spectrum, we convert the cylindrical 
geometry into a toroidal one, where periodic boundary conditions 
are imposed along 
the $x$ and $y$ directions. For rational $\alpha = p/q$, 
we write the spin-dependent Harper's Hamiltonian 
as a $2q \times 2q$ matrix in momentum $(k_x, k_y)$ space 
\begin{eqnarray}
\label{eqn:hamiltonian-matrix-toroidal-geometry}
{\bf H}({k_x, k_y}) 
= 
\left(
\begin{array}{c c}
    {\bf H}_{++} & {\bf H}_{+-}\\
    {\bf H}_{-+} & {\bf H}_{--}
\end{array}
\right)
\end{eqnarray}
by taking advantage of the magnetic translation group to define 
$q \times q$ block matrices ${\bf H}_{ss^\prime}$, where 
$\{s,s^\prime\} = \{ \pm, \pm \}$ label spin states 
$\vert \uparrow \rangle = \vert + \rangle$ 
and $\vert \downarrow \rangle = \vert - \rangle$. 
The spin-diagonal $q \times q$ block matrices 
${\bf H}_{ss}/t$ are 
\begin{eqnarray}
\left(
\begin{array}{c c c c c}
    \Gamma_1 & -e^{i k_{x s}a} & 0 & \dots  & -e^{-i k_{x s}a} \\
   -e^{-i{k}_{x s}a} & \Gamma_2 & -e^{i k_{x s}a} & \dots  & 0 \\
    \vdots & \vdots & \vdots & \ddots & \vdots \\
    -e^{i k_{x s}a} & 0 & \dots & - e^{-i k_{x s} a}  & \Gamma_q
\end{array}
\right),
\nonumber
\end{eqnarray}
where ${k}_{x s} = k_x -s k_T$ describes the spin-dependent 
momentum transfer along the $x$ direction. The kinetic energy terms are 
$\Gamma_m = -2\cos(k_ya - 2\pi\alpha m)$, with the magnetic flux ratio
being $\alpha = p/q$, and with $m$ taking values $(1, ..., q)$. 
The spin-off-diagonal $q \times q$ block matrices are 
\begin{eqnarray}
\frac{{\bf H}_{s \bar s}}{t} 
= 
\left(
\begin{array}{c c c c c}
    -h_x & 0 & 0 & 0  & 0 \\
    0 & -h_x & 0 & 0  & 0 \\
    \vdots & \vdots & \vdots & \ddots & \vdots \\
    0 & 0 & \dots & 0  & -h_x
\end{array}
\right),
\end{eqnarray}
describing spin-flip processes created by the independently 
tunable Zeeman field $h_x$, with $\bar s = -s$.
The resulting energy spectrum is essentially identical to open 
boundary problem with cylindrical geometry shown in Fig.~\ref{fig:two}, 
except that edge-state energy bands are not present because of the 
compactification to the toroidal geometry with periodic 
boundary conditions along the $x$ and
$y$ directions, that is, in this case there are no edges.

Next, we analyse the Chern spectrum that emerges for arbitrary spin-orbit
coupling and Zeeman fields and fixed flux ratio $\alpha = p/q$. 
The energy
spectrum associated with the Hamiltonian ${\bf H} (k_x, k_y)$ in
Eq.~(\ref{eqn:hamiltonian-matrix-toroidal-geometry}) has $2q$ 
spin-magnetic bands $E_{m_\sigma} ({\bf k})$ 
labeled by a magnetic band number $m_\sigma$ with generalized spin
index $\sigma$. The minimum number of gaps is $q-1$ when the bands 
are doubly degenerate and the maximum number of gaps is $2q - 1$ 
when there are no degeneracies. In the absence of overlaping regions between 
the energy bands $E_{m_\sigma} ({\bf k})$, the Chern index 
for the $m_\sigma^{th}$ band with generalized spin index $\sigma$ 
is
\begin{equation}
\label{eqn:chern-index}
C_{m_\sigma}  
= 
\frac{1}{2\pi i}
\int_{\partial \Omega} 
d^2 {\bf k} 
F_{xy}^{(m_\sigma)} ({\bf k}),
\end{equation}
where the domain of integration $\partial \Omega$ corresponds to the
magnetic Brillouin zone, that is, $\partial \Omega_x = [-\pi, \pi]$
and $\partial \Omega_y = -[\pi/q, \pi/q]$.  The function
\begin{equation}
F_{xy}^{(m \sigma)} ({\bf k}) 
= 
\partial_x A_y^{(m_\sigma)} ({\bf k}) 
- 
\partial_y A_x^{(m_\sigma)} ({\bf k}),
\end{equation}
is the Berry curvature expressed in terms of the Berry connection
$
A_{j}^{(m_\sigma)}({\bf k}) 
= 
\langle u_{m_\sigma} ({\bf k}) \vert 
\partial_j
\vert u_{m_\sigma} ({\bf k}) \rangle
$ 
where $\vert u_{m_\sigma} ({\bf k}) \rangle$ are
the eigenstates of the Hamiltonian ${\bf H} (k_x, k_y)$ 
defined in Eq.~(\ref{eqn:hamiltonian-matrix-toroidal-geometry}).
In the limit of zero spin-orbit coupling $(k_T = 0)$ and zero 
Zeeman field
$(h_x = 0)$, the energy spectrum for flux ratio $\alpha = p/q$ 
has doubly-degenerate $q$ magnetic bands and $q-1$ gaps, such 
that the Chern index from Eq.~(\ref{eqn:chern-index}) reduces 
to the standard form found in the quantum Hall effect 
literature~\cite{thouless-1982, kohmoto-1985}. 

To compute the Chern index $C_{m_\sigma}$, we generalize a 
discretization method used in the quantum Hall 
problem~\cite{hatsugai-2005} 
with zero spin-orbit coupling $(k_T = 0)$ and zero Zeeman field 
$(h_x = 0$). We define the link function 
$$
L_{j}^{(m_\sigma)} ({\bf k})
= 
\frac{
\langle u_{m_\sigma} ({\bf k}) 
\vert u_{m_\sigma} ({\bf k} + \delta {\bf k}_j)
\rangle
}
{
\vert 
\langle u_{m_\sigma} ({\bf k}) 
\vert u_{m_\sigma} ({\bf k} + \delta {\bf k}_j)
\rangle
\vert
}
= 
e^{i\theta_{j}^{(m_\sigma)} ({\bf k})}
$$
and obtain the discretized Berry curvature as
\begin{equation}
\label{eqn:berry-curvature}
F_{xy}^{m_\sigma} ({\bf k}) 
=
\ln
\left[
\frac{ 
L_x^{m_\sigma}({\bf k}) L _y^{m_\sigma} ({\bf k} + \delta {\bf k}_x)
}
{
L_x^{m_\sigma} ({\bf k} + \delta {\bf k}_y) L_y^{m_\sigma} ({\bf k})
}
\right],
\end{equation}
which is a purely imaginary number defined in the range 
$
-\pi \le 
{\cal I} \left[ F_{xy}^{m_\sigma} ({\bf k}) \right] 
\le \pi.
$
The Chern index becomes 
\begin{equation}
C_{m_\sigma} 
=
\frac{1}{2\pi i } \sum_{\bf k} F_{xy}^{(m_\sigma)}({\bf k}).
\end{equation}
When the energy bands $E_{m_\sigma} ({\bf k})$ overlap, that is, 
there are residual degeneracies in momentum space, we need to 
redefine the link variable of the degenerate bundle with 
degeneracy $D$ via the multiplet 
$
\vert \psi_{m_\sigma}^{(D)} ({\bf k}) \rangle
= 
\left[ 
\vert u_{m_\sigma}^{(1)} ({\bf k})\rangle,  
\dots ,  
\vert u_{m_\sigma}^{(D)} ({\bf k})\rangle
\right] ,
$
leading to 
$$
L_{j}^{(m_\sigma)} ({\bf k})
= 
\frac{
{\rm Det} \langle \psi_{m_\sigma}^{(D)} ({\bf k}) 
\vert \psi_{m_\sigma}^{(D)} ({\bf k} + \delta {\bf k}_j)
\rangle
}
{
\vert 
{\rm Det}
\langle \psi_{m_\sigma}^{(D)} ({\bf k}) 
\vert \psi_{m_\sigma}^{(D)} ({\bf k} + \delta {\bf k}_j)
\rangle
\vert
}
= 
e^{i\theta_{j}^{(m_\sigma)} ({\bf k})}
$$
with these new definitions, the expression for the Berry 
curvature defined in Eq.~(\ref{eqn:berry-curvature}) remains 
valid when written in terms of the new link functions defined 
above. For two internal states and magnetic flux 
ratio $\alpha = p/q$, there is a maximum of $2q$ non-overlaping 
bands and a maximum of $2q$ Chern indices.

Chern indices are properties of bands  
$E_{m_\sigma} ({\bf k})$ or band bundles with degeneracy $D$ 
and are independent of the location of the chemical potential $\mu$.
However, Chern numbers are defined only within band gaps and their 
values dependent on which gap the chemical potential is located.
If the chemical potential $\mu$ is located in a band gap 
corresponding to filling factor 
$\nu = r/q$, then the Chern number at this value of $\mu$ is the 
sum of Chern indices of bands with energies $E < \mu$. 
\begin{equation}
C_r = \sum_{m_\sigma, E < \mu}^{\nu = r/q} C_{m_\sigma}.
\end{equation}
Furthermore, via the bulk-edge correspondence~\cite{hatsugai-1996},
the Chern number $C_r$ calculated from the toroidal geometry 
(bulk system without edges) measures the total chirality of 
edge states that are present in the gap for the 
cylindrical geometry.

\begin{figure} [tb]
\centering 
\epsfig{file=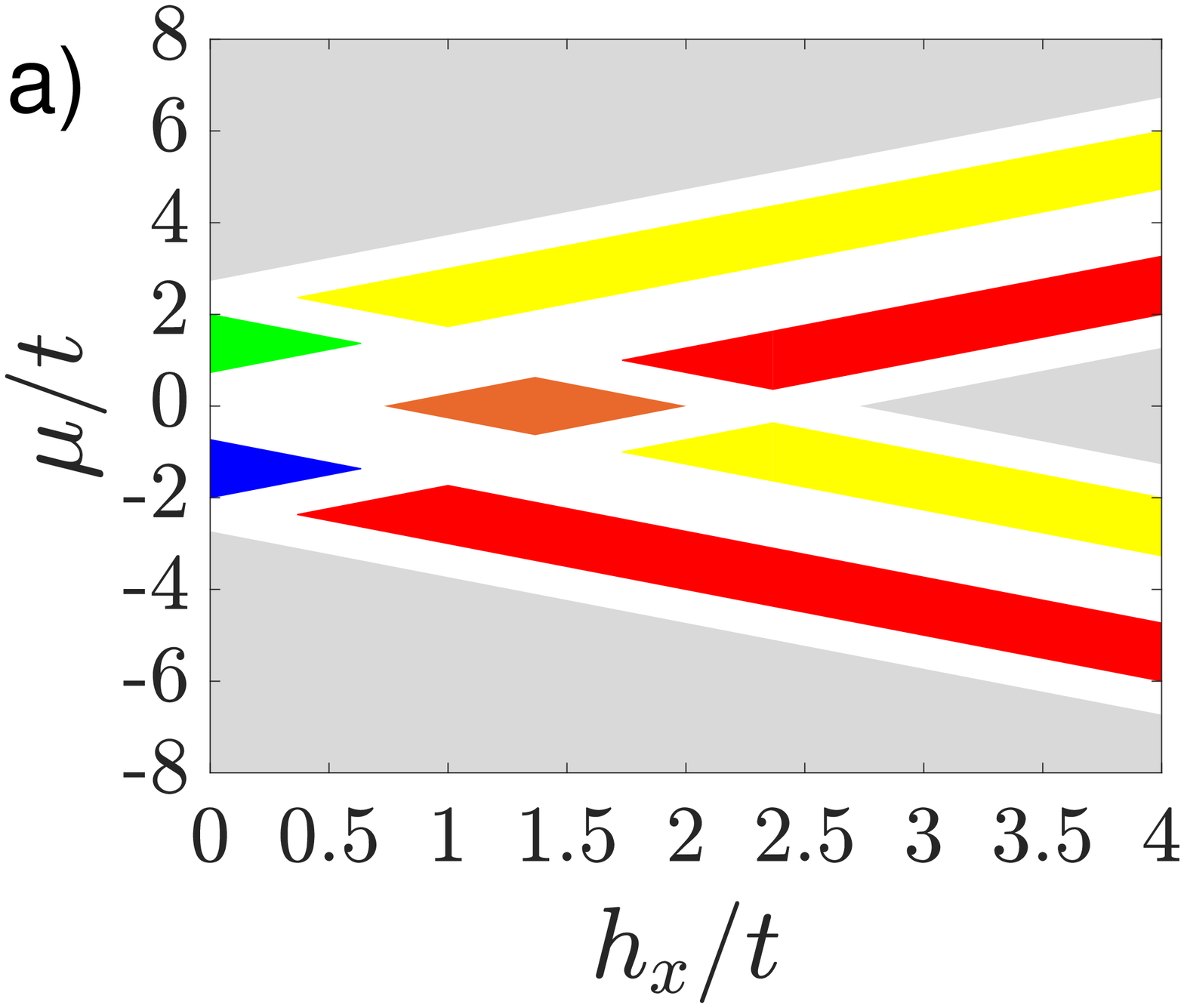,width=0.49 \linewidth}
\epsfig{file=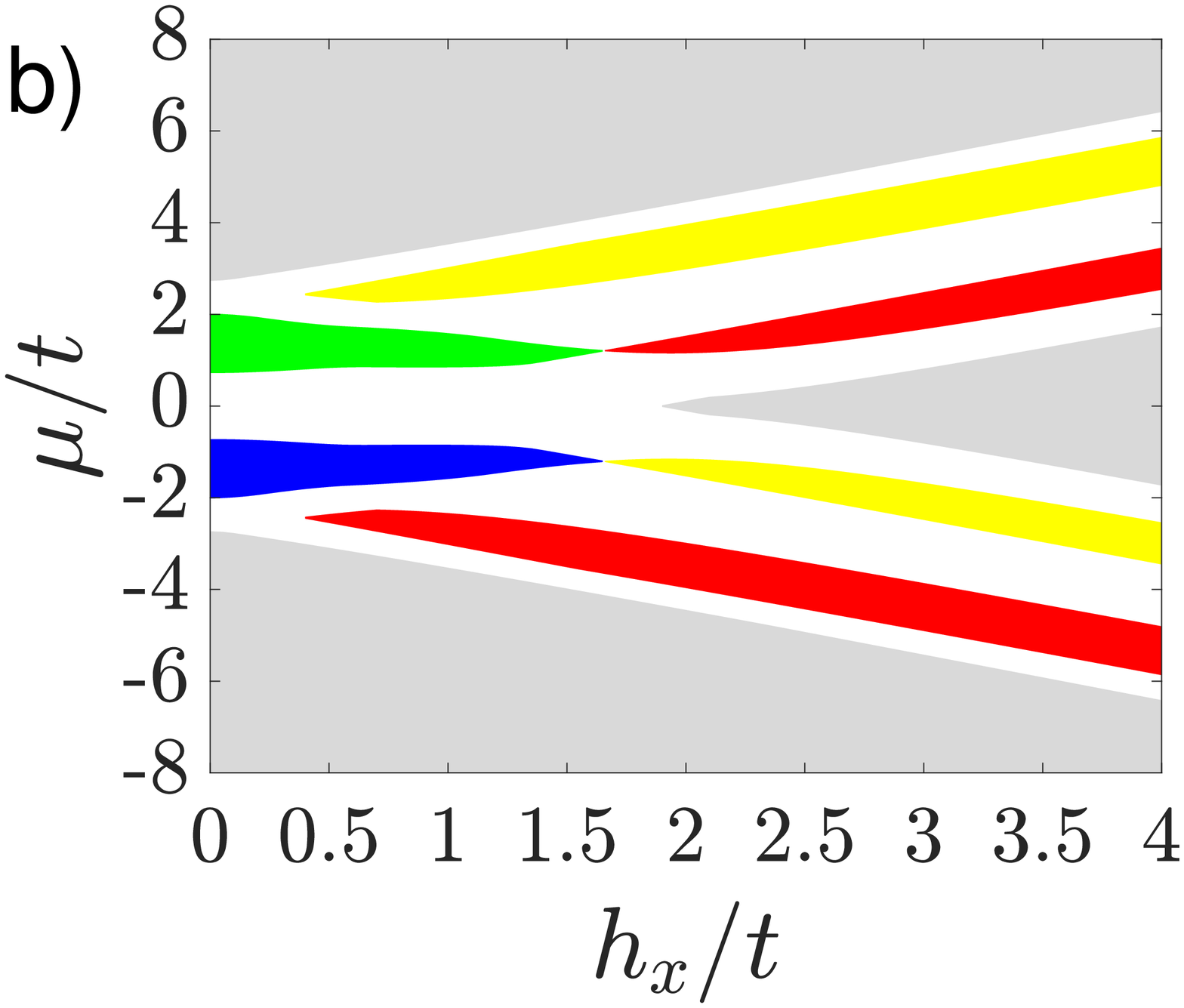,width=0.49 \linewidth}
\vskip 0.2cm
\epsfig{file=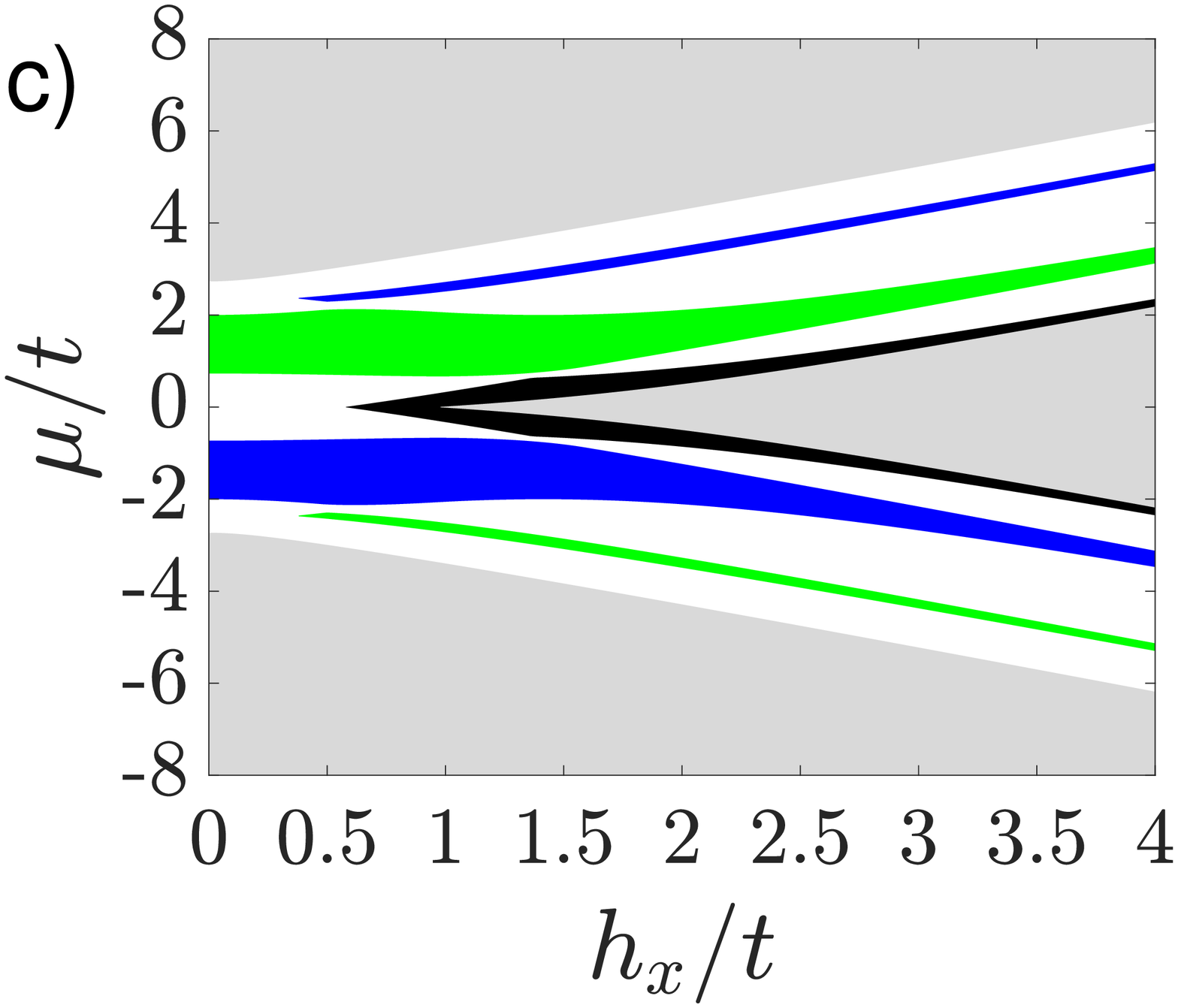,width=0.49 \linewidth}
\epsfig{file=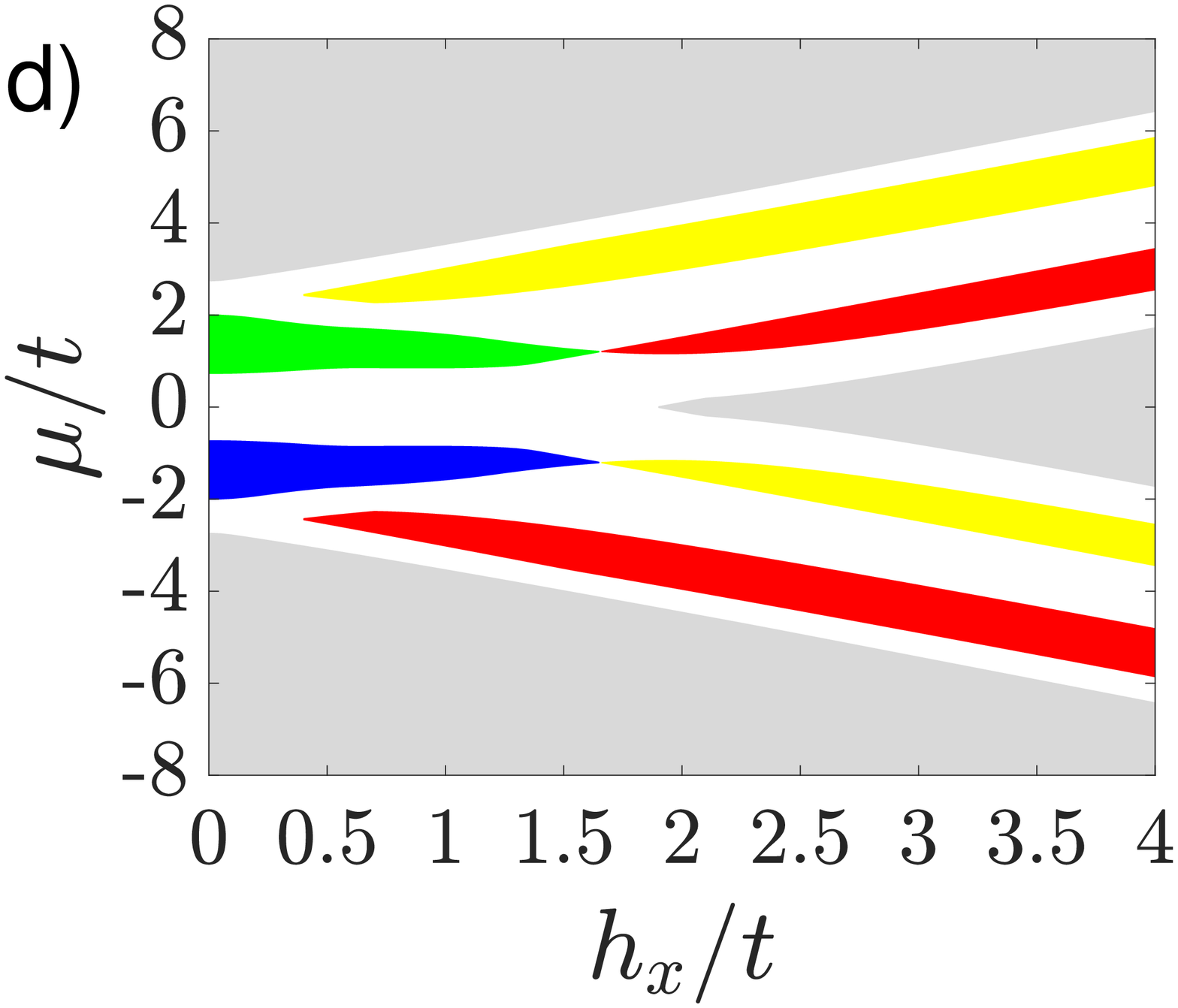,width=0.49 \linewidth}
\caption{ 
\label{fig:three}
(Color Online) 
Phase diagrams of chemical potential $\mu$ versus
Zeeman field $h_x$ and the associated Chern number spectrum
are shown for spin-orbit coupling parameters: a) $k_T a = 0$, 
b) $k_T a = \pi/4$ , c) $k_T a = \pi/2$ , and
d) $k_T a = 3\pi/4$. The white regions correspond to 
conducting phases and the non-white regions of different colors
correspond to insulating phases. The values of the Chern numbers
for each colored region are +2 (blue), +1 (red), 
0 (gray, black, orange), -1 (yellow), -2 (green). The gray regions are 
topologically trivial with no edge states, the black regions are
topologically trivial with non-chiral edge states, and the orange
regions are topologically non-trivial having two edge states with 
opposite chirality characteristic of quantum spin-Hall (QSH) phases .   
}
\end{figure}

{\it Phase Diagrams:} 
In Fig.~\ref{fig:three}, we show the phase diagram of chemical potential
$\mu$ versus Zeeman field $h_x$ for fixed value of the 
magnetic flux ratio $\alpha = 1/3$ and four values of 
the spin-orbit parameter:
a) $k_T a = 0$, b) $k_T a = \pi/4$, c) $k_T a = \pi/2$, and 
d) $k_T a = 3 \pi/4$. The white regions indicate
conducting phases, while the other colors indicate insulating phases
with Chern numbers $+2$ (blue), $+1$ (red), $0$ (gray, black, orange), 
$-1$ (yellow), $-2$ (green). The gray regions are 
topologically trivial with no edge states, the black regions are 
topologically trivial with non-chiral edge states, while the orange
region is topologically non-trivial having two edge states with
opposite chirality characteristic of a quantum spin-Hall (QSH) phase.  
For fixed magnetic flux $\alpha = 1/3$, the number of insulating phases 
between conducting regions increases from two $(2)$ at low Zeeman fields 
to five $(5)$ at high Zeeman fields. This does not occur 
in electronic systems, 
because the Zeeman field $h_x$ cannot be tuned 
independently of the magnetic ratio $\alpha$, 
and typically has very small values in comparison to 
the hoping parameter $t$, such that $h_x/t \ll 1$. 
However, for ultra-cold fermions, $h_x$ is a 
synthetic field that can be tuned independently of the magnetic ratio 
$\alpha$ and can attain high values in comparison to $t$. 

The Chern number spectrum is odd under inversion through 
filling factor $\nu = 1$ (around $\mu = 0$), 
as can be seen in all panels of Fig.~\ref{fig:three}.
Furthermore, the Chern number spectrum is even under inversion 
through $k_T a = \pi/2$, therefore it is the same for 
$k_T a = \pi/4$ and $k_T = 3\pi/4$, as seen in Fig.~{\ref{fig:three}b 
and Fig.~{\ref{fig:three}d. The lower and upper gray regions 
in all panels of Fig.~\ref{fig:three} 
correspond to trivial insulating phases with filling factors $\nu = 0$ 
and $\nu = 2$, respectively. In all insulating regions the system 
is incompressible, that is, $d\nu/d\mu = 0$.

In Fig.~\ref{fig:three}a, there is one insulating phase with 
filling factor $\nu = 1/3$ (lower red region), two insulating phases 
with $\nu = 2/3$ (blue and lower yellow region), two insulating phases
with $\nu = 1$ (orange and central gray region), two insulating phases
with $\nu = 4/3$ (green and upper red region), and one insulating phase
with $\nu = 5/3$ (upper yellow region). The most interesting feature 
of Fig.~\ref{fig:three}a is the orange region around $\mu = 0$ with 
filling factor $\nu = 1$, which exhibits the QSH effect. 
In Fig.~\ref{fig:three}b essentially the same phases of 
Fig.~\ref{fig:three}a are present, except for the orange region which 
disappears, because the spin-orbit parameter $k_T a$ is too large 
to preserve edge states with opposite chirality. The most interesting 
feature 
of Fig.~\ref{fig:three}b is the direct topological quantum 
phase transitions between the blue $(C_2 = +2)$ and an yellow  
$(C_2 = -1)$ regions at filling factor $\nu = 2/3$ 
and between the green $(C_4 = - 2)$ and red $(C_4 = +1)$ regions
at $\nu = 4/3$, where two chiral edge states disappear as the gap 
closes and a single chiral edge of opposite chirality emerges as 
the gap reopens. 
In Fig.~\ref{fig:three}c, where $k_T a = \pi/2$, there are only 
insulating phases with even Chern numbers: the
lower green region $(C_1 = -2)$ at $\nu = 1/3$, 
the lower blue region $(C_2 = +2)$ at $\nu = 2/3$, 
the central gray and black regions $(C_3 = 0)$ at $\nu = 1$, 
the upper green region $(C_4 = -2)$ at $\nu = 4/3$,
and the upper blue region $(C_5 = +2)$ at $\nu = 5/3$.
In Fig.~\ref{fig:three}d, where $k_T a  = 3\pi/4$, 
the phases shown are identical to those of Fig.~\ref{fig:three}b, 
where $k_T a = \pi/4$, because the Chern spectrum is even 
under inversion through $k_T a = \pi/2$.

The number and type of insulating states discussed here are 
very different from those of the integer quantum Hall effect found 
in semiconductor physics, when $h_x = 0$ and $k_T = 0$. 
For $\alpha = 1/3$, there are only two possibilities 
for the Chern number: $C_2 = +2$ at $\nu = 2/3$ and 
$C_4 = -2$ at $\nu = 4/3$, when $h_x = 0$ and $k_T = 0$. 
However, as $h_x$ is varied for 
fixed $k_T$ new topological insulating phases emerge with 
additional Chern numbers and filling factors.
For a fixed magnetic
ratio $\alpha = p/q$, the index $r$ labels gaps in the energy spectrum 
$E_{m_\sigma} ({\bf k})$ and is related to the integers $p$ and $q$ 
via the Diophantine equation 
$r = q S_r + p C_r$, where $S_r$ is a supplementary 
topological invariant and $C_r$ is the Chern number. This 
relation can be rewritten in terms of the filling factor 
$\nu = r/q$ and the magnetic ratio $\alpha = p/q$ as 
\begin{equation}
\label{eqn:diophantine}
\nu = S_r + \alpha C_r.
\end{equation}
Notice that $r$ can take a maximum value of $2q$, when $\nu = 2$. 
For $\alpha = 1/3$, incompressible phases with filling factors 
$\nu = 0$ $(r = 0)$ and $\nu = 2$ $(r = 6)$ have 
Chern numbers $C_0$ and $C_6$ trivially equal to zero, 
therefore $S_r$ can only take non-negative integer values $0, 1, 2$,
as $r$ varies from $0$ to $6$. 
The relation shown in Eq.~(\ref{eqn:diophantine}) generalizes 
the gap-labeling theorem~\cite{wannier-1978, claro-1979} used 
in the context of the integer quantum-Hall effect, because 
the topological quantum numbers
$(S_r, C_r)$ change not only as a function of the 
magnetic ratio $\alpha$, 
but also as a function of the Zeeman field $h_x/t$ and spin-orbit 
parameter $k_T a$.

\begin{figure} [tb]
\centering 
\epsfig{file=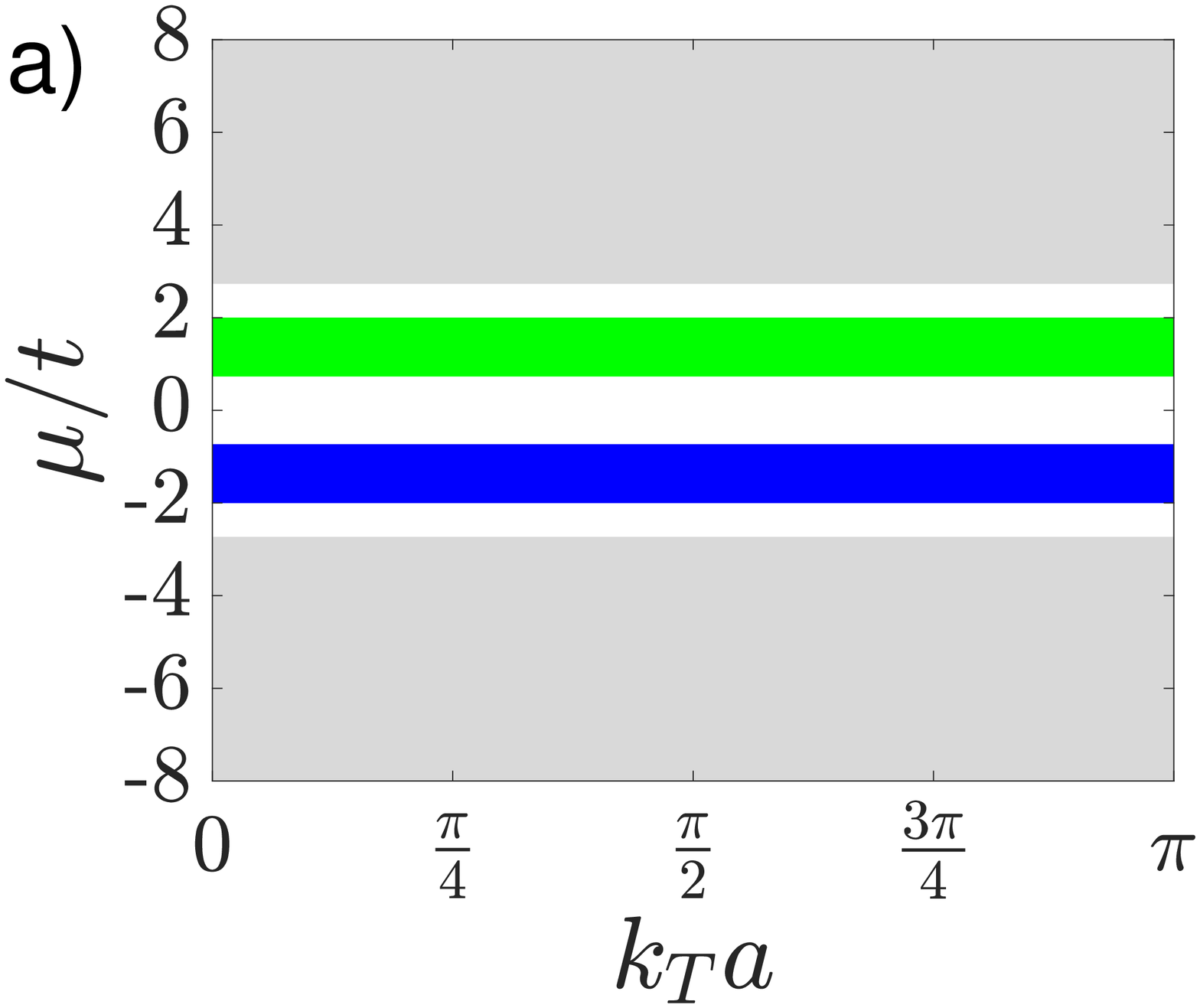,width=0.49 \linewidth}
\epsfig{file=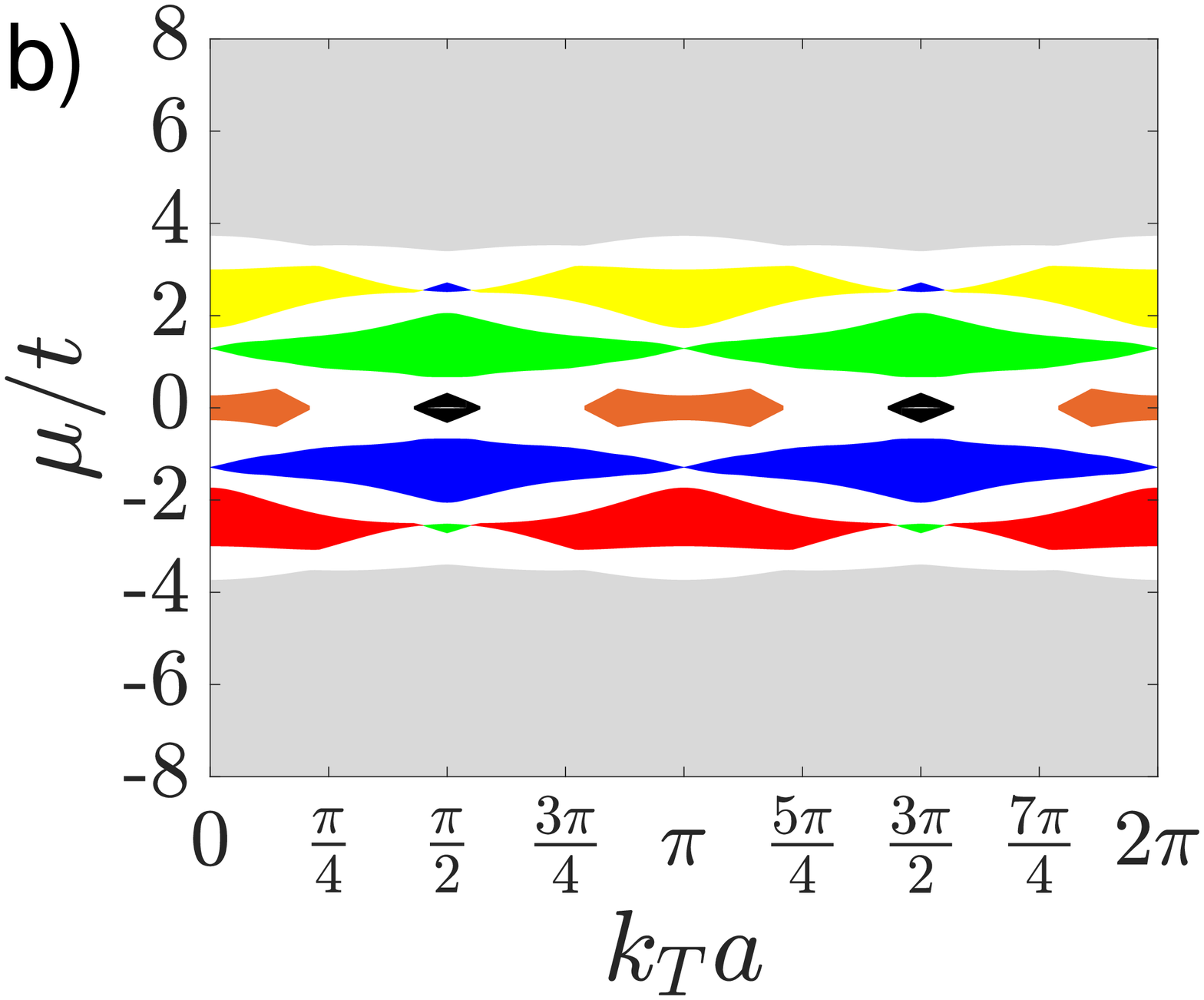,width=0.49 \linewidth}
\vskip 0.2cm
\epsfig{file=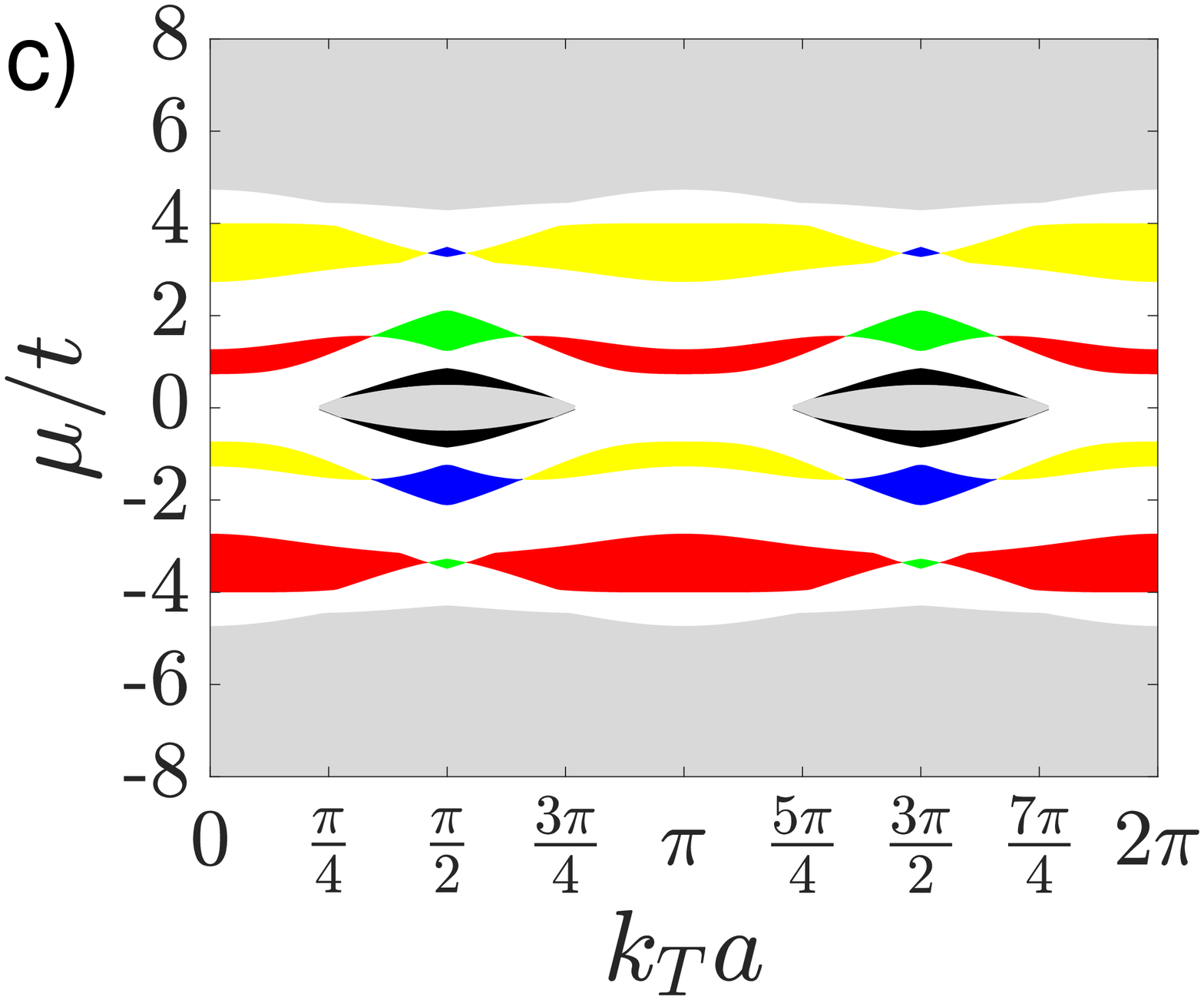,width=0.49 \linewidth}
\epsfig{file=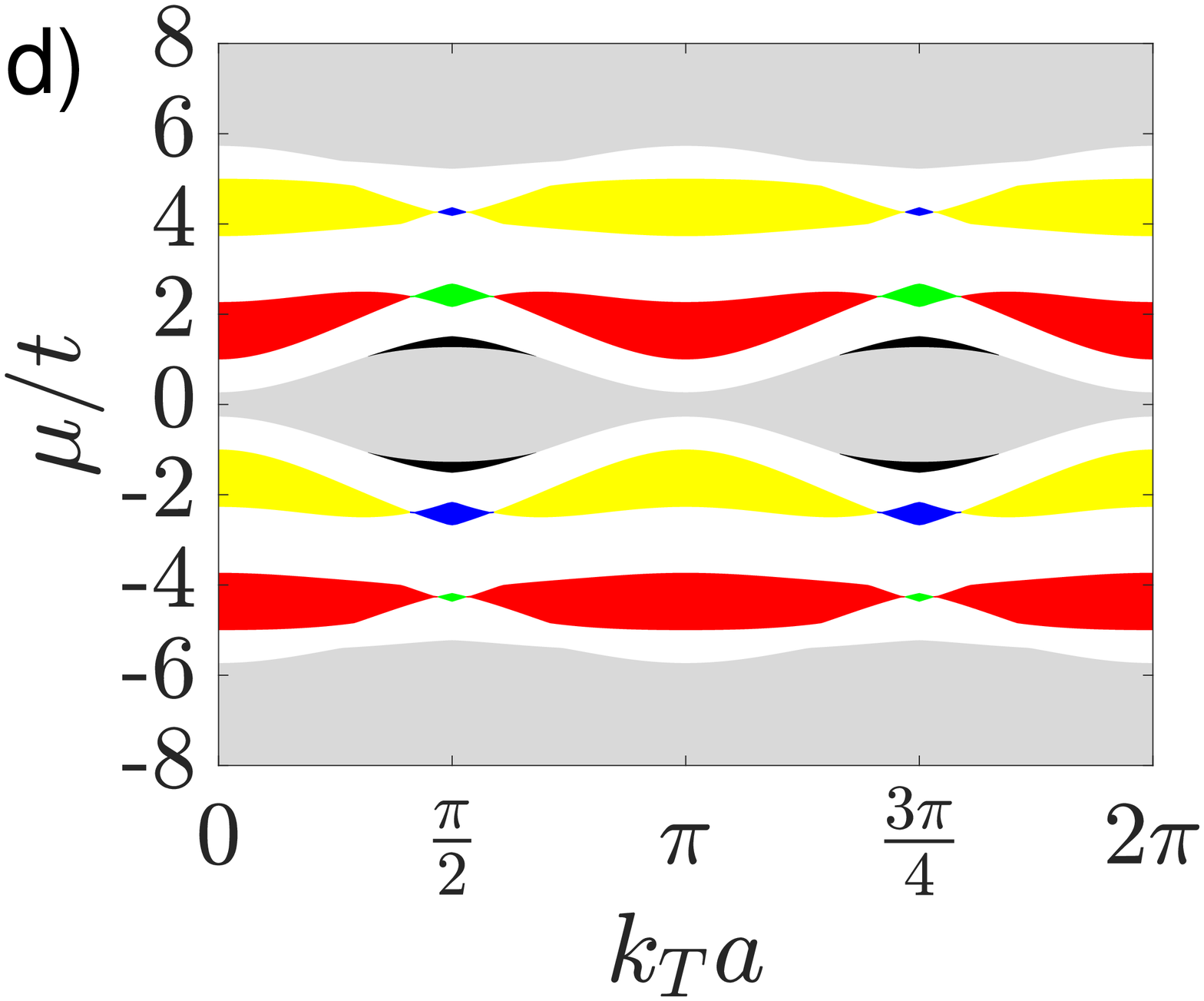,width=0.49 \linewidth}
\caption{ 
\label{fig:four}
(Color Online) 
Chemical potential $\mu/t$ versus spin-orbit parameter $k_T a$ for
flux ratio $\alpha = 1/3$ and Zeeman fields: 
a) $h_x/t = 0$,
b) $h_x/t = 1$,
c) $h_x/t = 2$,
d) $h_x/t = 3$.
The color code for the Chern numbers is the same as in
Fig.~\ref{fig:three}.
}
\end{figure}

In Fig.~\ref{fig:four}, we show the phase diagram of chemical 
potential $\mu/t$ versus spin-orbit parameter $k_T a$ 
illustrating all the insulating phases for 
magnetic ratio $\alpha = 1/3$ and changing Zeeman fields: 
a) $h_x/t = 0$, b) $h_x/t = 1$, c) $h_x/t = 2$ and d) $h_x/t = 3$. 
The color code for insulating phases is the same used in 
Fig.~\ref{fig:three}. Notice that the Chern spectrum in periodic 
in $k_T a$ with period equal to $\pi$, and that topological quantum 
phase transitions between insulators with different Chern numbers  
occur as $k_T a$ is changed for fixed $h_x/t$. The lower and upper gray 
regions correspond to $\nu = 0$ with $(S_0, C_0) = (0,0)$, and
$\nu = 2$ with $(S_6, C_6) = (2,0)$.

In Fig.~\ref{fig:four}a, where $h_x/t = 0$, 
there are only two topological insulating phases. The first one is the 
blue region at $\nu = 2/3$ with $(S_2, C_2) = (0, +2)$, and the second is 
the green region at $\nu = 4/3$, with $(S_4, C_4) = (2, -2)$. 
Because of the spin-gauge symmetry at $h_x/t = 0$, 
the values of $(S_r, C_r)$ are independent of the spin-orbit coupling
parameter $k_T a$. 
In Fig.~\ref{fig:four}b, where $h_x/t = 1$, more topological
insulating phases emerge at additional filling factors. 
At $\nu = 1/3$ there is a red region with 
$(S_1, C_1) = (0, +1)$ and a green region with
$(S_1, C_1) = (1, -2)$. At $\nu = 2/3$ there is a blue region with 
$(S_2, C_2) = (0, +2)$. At $\nu = 1$ there are orange and black regions 
with
$(S_3, C_3) = (1, 0)$.  At $\nu = 4/3$ there is a green region with
$(S_4, C_4) = (2, -2)$. At $\nu = 5/3$ there is a yellow region with
$(S_5, C_5) = (2, -1)$ and a blue region with
$(S_5, C_5) = (1, +2)$. 
Similar topological indexing can be done
for Figs.~\ref{fig:four}c and~\ref{fig:four}d, with the most important 
differences from Fig~\ref{fig:four}b being the emergence of
yellow regions at $\nu = 2/3$ with $(S_2, C_2) = (1, -1)$, 
gray regions at $\nu = 1$ with $(S_3, C_3) = (1,0)$,
and red regions at $\nu = 4/3$ with $(S_4, C_4) = (1, +1)$. 
Notice that the topologically non-trivial orange region (QSH insulator) 
of Fig.~\ref{fig:four}b disappears at larger Zeeman fields 
in Fig.~\ref{fig:four}c ($h_x/t = 2$) and
Fig.~\ref{fig:four}d ($h_x/t = 3$), as particle-like 
and hole-like magnetic bands no longer overlap, such that edge states 
with opposite chirality disappear leaving either non-chiral edge states
or no edge states at all.

Finally, notice a staircase structure of the filling factor $\nu$ 
versus chemical potential $\mu$ both in Fig.~\ref{fig:three} 
for fixed Zeeman field $h_x/t$ and in Fig.~\ref{fig:four} 
for fixed spin-orbit coupling $k_T a$. The steps of this staircase 
occur at values of $\nu$ given by the Diophantine relation in 
Eq.~(\ref{eqn:diophantine}).

{\it Conclusions:}
We have discussed the Chern number spectrum of ultra-cold fermions in 
square optical lattices as a function artificial magnetic, Zeeman and 
spin-orbit fields that can be tuned independently. 
As an specific example, we obtained phase diagrams of chemical potential 
versus Zeeman and spin-orbit fields for fixed magnetic flux ratio 
$\alpha = 1/3$. 
We showed that Chern numbers are dramatically modified when 
Zeeman and spin-orbit fields are changed for fixed magnetic field, 
and that topological quantum phase transitions between different
insulating states are induced by Zeeman and spin-orbit fields at fixed 
filling factor. Lastly, we obtained a staircase structure in the filling
factor versus chemical potential for changing Zeeman and spin-orbit 
fields, showing the existence of incompressible states at rational 
filling factors obtained from a generalized Diophantine equation.  

One of us (C.A.R.S.d.M.) would like to thank the support of 
the Galileo Galilei Institute for Theoretical Physics via a Simons
Fellowship, and of the International Institute of Physics via the 
Visitor's Program.


\end{document}